\documentclass{article}
\usepackage{graphicx} 
\usepackage[margin=1in]{geometry}
\usepackage{url}
\usepackage{tcolorbox}
\usepackage{caption}
\usepackage{newfloat}
\usepackage{amsmath}
\usepackage{longtable}
\usepackage[colorlinks=true]{hyperref}

\title{Generative AI in Live Operations: Evidence of Productivity Gains in Cybersecurity and Endpoint Management}
\author{James Bono, Justin Grana, Kleanthis Karakolios, \\ 
Pruthvi Hanumanthapura Ramakrishna, and Ankit Srivastava  \\ \\ Microsoft Corporation }
\date{March 2025}

\begin{document}

\maketitle

\begin{abstract}
We measure the association between generative AI (GAI) tool adoption and four metrics spanning security operations, information protection, and endpoint management: 1) number of security alerts per incident,  2) probability of security incident reopenings,  3) time to classify a data loss prevention alert, and 4) time to resolve device policy conflicts.  We find that GAI is associated with robust and statistically and practically significant improvements in the four metrics.    Although unobserved confounders inhibit causal identification, these results are among the first to use observational data from live operations to investigate the relationship between GAI adoption and security operations, data loss prevention, and device policy management.
\end{abstract}

\section{Introduction and Background}
Recent developments in generative artificial intelligence (GAI) have raised questions about its productivity effects. Although GAI-based productivity improvements have important implications for the labor market \cite{ailit1,ailit2},
its implications in cybersecurity and information technology governance extend beyond labor cost savings and also increase the potential to reduce costly attacks.  Despite its broad implications, there is scant evidence detailing the potential benefits of GAI tools for security and information workers in live operations.

In an effort to continue gathering this evidence, this paper analyzes the impact of GAI adoption on four crucial security and management metrics. Specifically, we leverage the tools of causal inference (difference-in-differences, propensity score matching, and two-way fixed effects regressions) to measure the association between GAI adoption and four metrics: 1) number of security alerts per incident;  2) probability of a security incident reopening;  3) mean time to classify (MTTC) a data loss prevention (DLP) alert; and 4) mean time to resolve (MTTR) device policy conflicts.  In our study, we say that an organization adopted GAI tools if they use Microsoft Security Copilot (``Copilot''), which ``combines a specialized language model with security-specific capabilities''\cite{cfs}.

 Our results show that GAI adoption is associated with robust and statistically significant productivity improvements for all four metrics.  Specifically, GAI adoption is associated with   
 1) a 22.88\% decrease in the number of alerts per incident; 2)  a 68.44\% decrease in the probability of an incident being reopened; 3) an 18.38\% reduction in time to classify a DLP alert; and 4) a 54.34\% reduction in time to resolve a device policy conflict.

\subsection{Related Work}
Our work is an extension of \cite{mttrlive}, which examined how Copilot adoption is associated with security incident MTTR using data from live operations.  From a broader productivity perspective, our work sits adjacent to recent studies that examine the impact of GAI on worker productivity utilizing randomization in a laboratory or field setting \cite{defender1,git1,field1,exp1,leg1}.  While laboratory experiments provide solid causal identification, whether they generalize to real-world settings is a matter of judgment about the experimental design. Field experiments make a similar trade-off to lab studies. Random assignment gives them strong causal identification, but because they are often limited to a small set of organizations, the measured effects may not generalize more broadly. 

Like \cite{mttrlive}, our work makes the opposite trade-off; we examine telemetry data from live operations from large numbers of heterogeneous organizations, so our results are likely to generalize to live operations for a broad range of organizations. However, our causal claims are weaker than laboratory and field experiments because of our inability to control for selection into treatment. Our work is more related to the limited studies that use observational data \cite{genai_live,git2} in live operations to estimate the impact of GAI on productivity.  As is typical in these studies, unconfoundedness cannot be guaranteed, and thus our results provide evidence for ---  though do not causally identify --- GAI's impact on security worker productivity.  Despite their different shortcomings, studies have yielded surprisingly consistent estimates of GAI's productivity effects across domains (see Table \ref{tab:sum} for specific estimates) with our 18-30\% estimates being similarly consistent.

\begin{table}[h]
\centering
\begin{tabular}{|p{1cm}|p{7cm}|p{7cm}|} \hline
\textbf{Source} & \textbf{Domain} & \textbf{Estimated Productivity Gain}  \\ \hline 
\cite{genai_live} & Customer technical support & 34\% reduction in task completion time for novices\\ \hline 
\cite{defender1} & Security incident laboratory experiment & 23\% decrease in task completion time \\\hline 
\cite{git1} & Laboratory experiment implementing HTTP server in Javascript & 55.8\%  decrease in time to completion\\ \hline 
\cite{field1} & Field study of software development tasks & 26.08\% increase in tasks completed \\ \hline
\cite{itrct} & Laboratory study of IT Admins & 34.53\% accuracy improvement and 30.69\% reduction in task completion time \\ \hline
\end{tabular}
\caption{Estimated Productivity Gains from Generative AI}
\label{tab:sum}
\end{table}

From a security perspective, our study builds on research about automation of security tasks. The literature on this topic finds that risk mitigation is essential with the growth of security breaches, especially because so many security vulnerabilities operate at the gap between how systems are supposed to operate and how they actually operate \cite{morgan}. Although many aspects of cybersecurity currently rely on human subject matter experts \cite{costa}, researchers point out the possibility of automating error-prone and time-consuming security work. Machine learning techniques show particular promise in intelligently analyzing cybersecurity data \cite{sarker}. Natural language processing, knowledge representation and reasoning, and rule-based expert systems modeling can also support AI-driven cybersecurity \cite{sh}.

\subsection{Domain Background}

\paragraph{What is Security Event Management?} Security event triage and response is a central function of an organization's cybersecurity operations \cite{event}.  Broadly, computer network activity generates telemetry (logs, signatures, etc.).  That telemetry is then collated and processed by tools such as security information and event management (SIEM) and extended detection and response (XDR) solutions. These solutions often generate security \textbf{alerts} based on logical rules or statistics and machine learning models.  Ultimately, SIEM and XDR solutions aggregate telemetry into a set of distinct units, representing holistic depictions of suspicious activity, for human analysts to investigate.  These aggregations of alerts are commonly called \textbf{incidents}.   

Teams of analysts triage and resolve alerts and incidents.  Some may be resolved as a ``false positive'' where the suspicious behavior is determined to be benign.  Other incidents may require more intensive remediation and response such as altering users' privileges, disconnecting compromised systems, removing malicious files, and applying patches. Once an analyst resolves an alert or incident, they can proceed to the next  in their SIEM or XDR solution. 
However, the incident arrival rate far exceeds what a typical SOC can effectively triage. Recent estimates suggest that as much as 67\% of security incidents go unresolved \cite{vect}.

A DLP alert is a specific type of alert.   Administrators can set policies that manage data use. When those policies are violated, automated responses are triggered and administrators are notified with an alert.  For example, administrators can set a rule that blocks copying sensitive documents to a USB drive. If a user then tries to copy a sensitive document to a USB drive, the action is blocked and the administrator is notified.  Although rule-based, DLP alerts can still be false positives.  For example, a DLP policy can restrict address sharing for privacy reasons but may incorrectly flag an email exchange that discusses Pulitzer Prize-winning novel ``The Road'' by Cormac McCarthy.  Therefore, administrators must still classify alerts as true positives or false positives. 

\paragraph{What is Device Policy Management?}
Device Policy Management (DPM) is essential for maintaining secure and compliant devices within an organization. Administrators configure policies that determine device settings related to security and compliance. For example, Microsoft Intune can enforce a policy requiring all devices to have strong passwords, ensuring sensitive company data remains protected. One challenge in DPM arises due to a common situation in which multiple policies are applied to the same devices and the policy settings come into conflict, e.g., policies enforcing different password requirements. With many policies, settings, and devices, it can be challenging to efficiently identify the points of conflict and remediate them, which can delay the deployment of important policies and the protection they provide, as well as consume valuable IT admin resources. 

\begin{figure}[ht!]
   \centering
   \includegraphics[width=0.9\linewidth]{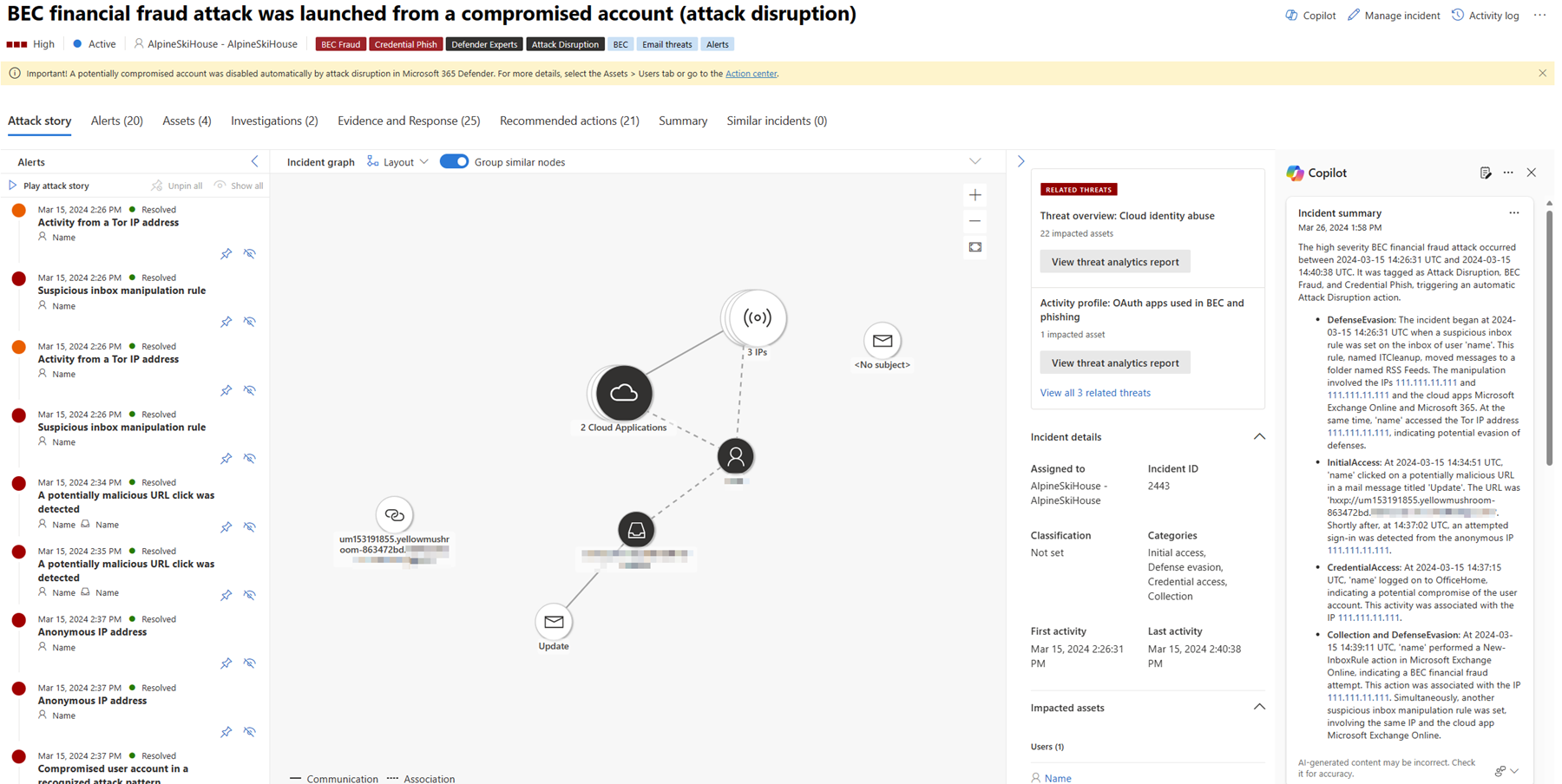}
   \caption{ An example of incident summarization for a Business Email Compromise Incident}
   \label{fig:sc_ex}
\end{figure}

\paragraph{What is Copilot?} Microsoft Copilot is a GAI tool designed to enhance the operational efficiency and effectiveness of security and IT professionals.  A key capability of Copilot is its ability to summarize complex data in a form that human analysts can quickly understand, an important function in security and IT domains where analysts make decisions based on rich data from a variety of sources.
The right panel of Figure \ref{fig:sc_ex} illustrates an example of incident summarization for a business email compromise security incident. Here, Copilot summarizes the alerts involved in the security incident, highlighting the most salient features for the analyst.

Although a full list of the product features is beyond the scope of this paper, we mention the following additional features to better illustrate the mechanisms by which Copilot impacts productivity:  

\begin{itemize}
    \item Guiding analysts to the appropriate response actions based on the incident data.
    \item Helping interpret malicious scripts associated with the incident.
    \item Creating scripts to query security logs based on natural language inputs.
    \item Retrieving relevant threat intelligence using natural language inputs.
    \item Summarizing device management policies.
\end{itemize}

\subsection{Security, Data Protection, and Device Management Productivity Metrics}
There are several metrics that quantify productivity in security operations, DLP, and DPM. Among them, the time it takes to perform a specific task provides the clearest relationship to productivity. For the security domain, previous work \cite{mttrlive} has already shown an association between Copilot adoption and reductions in MTTR. Here, we adopt task-duration metrics for our other domains: time to classify a DLP alert and time to resolve DPM conflicts. Naturally, if Copilot reduces the time to complete an essential task without sacrificing the quality of the task, productivity has increased.  

 Having already measured MTTR for security incidents, in this work we use two different metrics for the security domain.  The first metric --- number of security alerts per incident --- is similar to a time-based metric.  However, instead of considering wall-time, the metric captures how much ``activity'' has transpired before an incident is resolved.  Less activity, as measured by the number of alerts, suggests that incidents are resolved earlier in the kill-chain.  The second metric --- probability of incident reopening --- captures accuracy improvements.  The intuition is that, if an incident is marked resolved and then reopened as new information arrives, the initial incident resolution was incorrect and needs adjustment.  Therefore, a decrease in incident reopening rates represents more accurate initial decisions, thus improving productivity by making analysts more accurate.

\section{Alerts per Incident and Security Incident Reopenings}
\label{sec:nonmttr}

\subsection{Data and Method}
We collected system metadata generated by the Microsoft Defender XDR product spanning the period between August 18, 2024, and February 18, 2025. To construct our treatment group, we select organizations that have adopted Copilot and have at least 30 closed incidents before and after adopting Copilot, resulting in 378 organizations. To construct our control group, we use propensity score matching to select the organizations that are most similar to our treatment group. For the propensity score matching, we use logistic regression and regress whether an organization adopted Copilot on organization segment, industry, country, strategic category,
 and available seats per Microsoft security product. We then select matches by minimizing the sum of squared differences of propensity scores.  

Prior to modeling, we implement a targeted filtering approach to ensure data quality by removing anomalous data points that might distort our findings. Specifically, we remove incidents that had more than one thousand alerts as well as those that belong to organizations with an incident reopening rate above 0.8. This preprocessing step removes approximately 12\% of the data set while preserving the integrity of our causal inference framework. 

To measure the impact of Copilot on the number of alerts per incident, we estimate:

\begin{equation}
    N_{it} = \alpha + \eta_1  I[Treatment_i] + \sum_{\tau=1}^3\bigg[\gamma_{\tau} I[\tau=t] + \beta_{\tau} I[\tau=t]\times I[Treatment_i]\bigg] + \text{controls} 
\end{equation}
where
\begin{itemize}
\item $i$ indexes incidents.
\item $t$ indexes months after adoption.
\item $N_{it}$ is the number of alerts in incident $i$.
\item $I[Treatment_i]$ indicates if the incident originated from an organization that adopted Copilot.
\end{itemize}
The $\beta_t$ parameters capture the difference-in-differences between the number of alerts per incident of the treatment and control group $t$ months after adoption.  This model is what \cite{dyn_did} refers to as the ``dynamic specification'' of a two-way fixed effects model but removes the organization level fixed effects. Instead of controlling for organization-level effects, we control for incident-level effects because incident interactions are more likely to be incident-specific rather than organization-specific.  In this regression, we control for the product that triggered the alert as well as the alert category (exfiltration, initial access, etc.).

To measure the impact of Copilot on the probability of an incident reopening, we  estimate a similar model
\begin{equation}
    P(y_{it}=1) = \text{exp}  \bigg(\alpha + \eta_1 I[Treatment_i] + \sum_{\tau=1}^3\bigg[\gamma_{\tau} I[\tau=t] + \beta_{\tau} I[\tau=t]\times I[Treatment_i]\bigg] + \text{controls}  \bigg)
\end{equation}
where 
\begin{itemize}
\item $i$ indexes incidents.
\item $t$ indexes months after adoption.
\item $y_{it}=1$ if incident $i$ in time $t$ is reopened.
\item $I[Treatment_i]$ indicates if the incident originated from an organization that adopted Copilot.
\end{itemize}
The incident characteristics we use as part of the controls are the number of high-severity alerts in the incident and the total number of alerts in the incident.  
Ideally, we would also control for alert product and category in our logistic regression.  However, the alert products and categories are correlated with the number of alerts (distinct alert products and categories define the minimum number of alerts). Hence, we opt for the more compressed representation of alert characteristics and just use alert counts for the sake of parsimony.    We use cluster-robust standard errors at the organization level to conduct statistical inference in both models.

\subsection{Results}

Table \ref{tab:nalerts} gives the main result for the impact of GAI on the number of alerts per incident. In summary, we observe statistically significant reductions in the number of alerts per incident for the first and third month after adoption.  The lower estimate and lack of statistical significance in the second month is not concerning and likely due to anomalous events in the incident generation process.  This is supported by the fact that the estimate in the third month after adoption is similar in magnitude to the first month after adoption. Given the distribution of alerts per incident, this amounts to an estimated 22.88\% reduction in alerts per incident three months after adoption. 

\begin{table}
    \begin{center}
\begin{tabular}{|c|c|c|c|c|}
\hline
\textbf{Parameter} & \textbf{Estimate} & \textbf{Clustered S.E.} & \textbf{t-statistic} & \textbf{p-value} \\
\hline \hline
$\beta_1$ & -0.8051 & 0.2401 & -3.3527 & 0.000846 \\
\hline
$\beta_2$ & -0.3805 & 0.2512 & -1.5145 & 0.130383 \\
\hline
$\beta_3$ & -0.7408 & 0.2932 & -2.5262 & 0.011764 \\
\hline
\end{tabular}
\caption{Results for GAI's impact on Alerts per Incident. $R^2=0.014$}
    \label{tab:nalerts}
    \end{center}
\end{table}

The result of our regression on the probability of incident reopening is given in Table \ref{tab:reopen}.  Again, we see a statistically significant effect in the first and third months after adoption with the second month being marginally not significant ($p=0.11$). However, the effect in the third month is the most pronounced, suggesting that organizations are learning how to use the GAI tool.  

To understand the practical implications of our parameter estimate, we first collect incidents and their characteristics for adopters three months after adoption.  We then calculate the difference between the predicted probability of an incident reopening conditional on the organization being an adopter and the predicted probability conditional on the organization not adopting.  We normalize this difference by the predicted probability given an organization did not adopt to get the percentage reduction in probability of reopening for each incident.  Averaging this quantity across all incidents yields our estimate of an average 68.44\% decrease in the probability of an incident being reopened.  

\begin{table}[h!]
\centering
\begin{tabular}{|c|c|c|c|c|}
\hline
\textbf{Parameter} & \textbf{Coefficient} & \textbf{Clustered S.E.} & \textbf{t-statistic} & \textbf{p-value} \\
\hline \hline
$\beta_1$ & -0.399536 & 0.165074 & -2.420339 & 0.015506 \\
\hline
$\beta_2$ & -0.386737 & 0.244300 & -1.583041 & 0.113412 \\
\hline
$\beta_3$ & -1.159536 & 0.419890 & -2.761525 & 0.005753 \\
\hline
\end{tabular}
\caption{Results for GAI's impact on Probability of Reopening. Pseudo $R^2=0.014$}
\label{tab:reopen}
\end{table}

\subsection{Robustness}

To ensure the reliability of our findings for the reduction in alerts per incident, we conducted several robustness checks by varying key data filtering thresholds to handle outliers. Table \ref{tab:alerts_robust} presents these results with clustered standard errors by organization. For each specification, we report the estimated effect in the third month after adoption (Month 3+), the corresponding t-statistic, p-value, and the percentage reduction in alerts per incident relative to the baseline. Our base model demonstrates a statistically significant 22.88\% reduction in alerts per incident three months after Copilot adoption (p=0.012). When applying increasingly stringent data filtering criteria, the estimated effect magnitudes remain substantial, ranging from 13.12\% to 30.06\% reduction. The consistency of negative coefficients across all specifications supports our main finding that Copilot adoption reduces alert volume per incident, though statistical significance varies when accounting for within-organization correlation through clustered standard errors. The most stringent filtering approach (removing incidents with $>$1000 alerts and duration $<$1 minute) yields the largest estimated reduction of 30.06\%, with marginal statistical significance (p=0.101).

\begin{table}[h]
\centering
\resizebox{\textwidth}{!}{
\begin{tabular}{|l|c|c|c|c|c|}
\hline
\textbf{Specification} & \textbf{Month 3+ Effect} & \textbf{Clustered S.E.} & \textbf{t-statistic}  & \textbf{p-value} & \textbf{\% Reduction} \\
\hline
Base model (original) & -0.7408 & 0.2932 & -2.5266 & 0.012 & 22.88\% \\
\hline
\multicolumn{6}{|l|}{\textit{Alternative Data Filtering:}} \\
\hline
No alert filtering & -0.3582 & 0.397 & -0.9022  & 0.367 & 13.12\% \\
\hline
Filter $>500$ alerts & -0.5709 & 0.428 & -1.3339  & 0.182 & 24.29\% \\
\hline
Filter $>1000$ alerts, duration $<1$m & -0.7411 & 0.452 & -1.6397  & 0.101 & 30.06\% \\
\hline
Filter $>1000$, duration $<1$m \& $>24$h & -0.4797 & 0.412 & -1.1643  & 0.244 & 25.82\% \\
\hline
\end{tabular}
}
\caption{Robustness Checks for Number of Alerts per Incident Analysis}
\label{tab:alerts_robust}
\end{table}

Similarly, to assess the robustness of our findings regarding Copilot's effect on incident reopening rates, we conducted sensitivity analyses across different data filtering criteria and model specifications. Our base model, which excludes organizations with anomalously high reopening rates ($>$80\%), shows a substantial and statistically significant coefficient of -1.1596 (p=0.006), representing a 68.44\% reduction in reopening probability in the third month after Copilot adoption.

\begin{table}[h]
\centering
\begin{tabular}{|l|c|c|c|c|c|}
\hline
\textbf{Specification} & \textbf{Coefficient} & \textbf{Clustered S.E.} & \textbf{t-statistic} & \textbf{p-value} & \textbf{\% Reduction} \\
\hline
Base model (filter $>0.8$) & -1.1596 & 0.4199 & -2.762 & 0.006 & 68.44\% \\
\hline
\multicolumn{6}{|l|}{\textit{Alternative Data Filtering:}} \\
\hline
No filtering & -0.5467 & 0.3676 & -1.487 & 0.137 & 18.90\% \\
\hline
Filter reopening rate $>0.6$ & -1.0896 & 0.7246 & -1.504 & 0.133 & 70.26\% \\
\hline
\multicolumn{6}{|l|}{\textit{Alternative Model Specifications:}} \\
\hline
With Security Products & -1.1824 & 0.7958 & -1.486 & 0.137 & 71.98\% \\
\hline
With Category FE & -1.2258 & 0.8007 & -1.531 & 0.126 & 74.20\% \\
\hline
\end{tabular}
\caption{Robustness Checks for Probability of Incident Reopening}
\label{tab:reopen_robust}
\end{table}

Although the directional effect remains consistent across specifications, the results are less statistically robust, as the p-value for the treatment variable falls between $0.12$ and $0.14$.   Without filtering high-reopening organizations, the effect decreases to 18.90\% (p=0.137), indicating that extreme reopening patterns attenuate the treatment effect estimate. Alternative specifications show consistent effect sizes ranging from 71.98\% to 74.20\%. In all, while statistical significance is weaker than in our base model, it is reassuring that all of the estimated effect sizes are directionally the same and largely of similar magnitude.


\section{Data Loss Prevention Alerts}

\subsection{Data and Method}

\subsubsection{Data}
To estimate the association between GAI adoption and DLP productivity, we collect system metadata from 71 organizations that receive data loss prevention alerts.  Our data was collected from September 13, 2024 through March 13, 2025 and includes 262,718 alerts.  We only include alerts that are classified within 60 minutes, though we later show our results are robust to this assumption.  We also collect whether an alert was classified via a web portal or via an API.  The rationale for this is to control for alerts that are likely classified by a human analyst (web portal) or an automated alert classification pipeline that calls an API.  We call this the classification source. We denote an organization's GAI adoption date as the first date on which they use Microsoft Copilot to summarize a compliance alert.  We also capture whether the alert was classified as a true positive or a false positive.  We calculate the time to classification as the time an alert is first raised to the time it is classified.

As a technical aside, there are two types of compliance alerts, data-loss-protection and insider risk management.  Due to privacy policies, we cannot discern the type of alert summarized, so our adoption date may understate the true first date in which an organization summarized a data loss prevention alert.  This would lead to an attenuation of our estimate and thus our estimate on time savings should be considered a lower bound on productivity enhancements.

\subsubsection{Method}
To estimate the impact of GAI adoption on time to classification, we estimate a two-way fixed effects model with heterogeneous effects based on the classification (true positive or false positive) and the classification source (web portal or API).  We estimate:
\begin{equation}
\log(y_{ijt}) = A_i + B_t + \beta_0 T_{it} + \gamma_0 C_{ijt} + \gamma_1 S_{ijt} + + \gamma_2 S_{ijt}*C_{ijt} + \beta_1T_{it}*C_{ijt} + \beta_2 T_{it}*S_{ijt} + \beta_3 T_{it}*C_{ijt}*S_{ijt}
\end{equation}
where $y_{ijt}$ is alert $j$ for organization $i$ at time  $t$ (measured in weeks).  The variables $A_i$ and $B_t$ are organization and time fixed effects and $T_{it}$ indicates whether organization $i$ has adopted  Microsoft Copilot by time $t$.  $C_{ijt}$ is a dummy variable indicating whether alert $i,j,t$ is a true positive and $S_{i,j,t}$ indicates whether the alert was classified via a web portal.  We use the logarithm of time as the dependent variable to estimate the percent change in classification time as a result of adopting GAI.  The parameters of interest are $\beta_i, \; i=0...3$ where the calculated effect depends on the classification and classification source.  To be conservative, we cluster standard errors at the organization-week level, though our results are robust to alternative clustering choices, including at the organization level.

\subsection{Results}

\begin{figure}[h]
\centering
    \includegraphics[width=.8\textwidth]{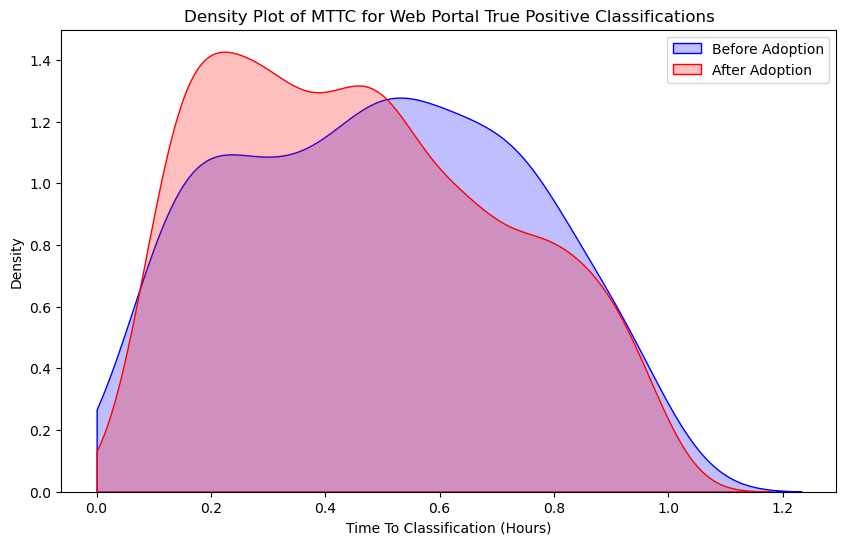}
    \caption{Empirical Distribution of MTTC}
    \label{fig:dlpres}
\end{figure}

Our results quantify the shift in distributions depicted in Figure \ref{fig:dlpres} while controlling for organization and time fixed effects. Table \ref{tab:DlpRes} presents the results for the main parameters of interest.  

\begin{table}
    \begin{center}
\begin{tabular}{|l|c|c|c|c|}
\hline
\textbf{Parameter (Variable)} & \textbf{Estimate} & \textbf{Clustered S.E.} & \textbf{t-statistic} &\textbf{p-value} \\ \hline \hline
$\beta_0 \; (T_{it})$  & -0.0054 & 0.01  & -0.55   & 0.582 \\ \hline
$\beta_1 \; (T_{it}*C_{ijt})$  & 0.0439  & 0.018 & 2.458   & 0.014 \\ \hline
$\beta_2 \; (T_{it}*S_{ijt})$  & 0.0013  & 0.12  & 0.011   & 0.991 \\ \hline
$\beta_3 \; (T_{it}*C_{ijt}*S_{ijt})$   & -0.243  & 0.122 & -1.986  & 0.047 \\ \hline
\end{tabular}
\end{center}
\caption{Results for GAI's impact on Data Loss Prevention. $R^2=0.479$}
\label{tab:DlpRes}
\end{table}

Table \ref{tab:DlpRes} indicates that the only practically meaningful reduction in MTTC is for true positives classified in the web portal ($\beta_3$).  However, the full treatment effect of GAI on true positives classified in the web portal is given by $\sum_{i=0}^3\beta_i=-0.2031$.  Using the cluster-robust standard errors to test $\sum_{i=0}^3\beta_i=0$ yields an $F$-statistic of 4.93 $(p=0.0267)$ allowing us to reject the hypothesis at the 5\% significance level. Since our model is a log-linear regression, we compute $\exp(\sum_{i=0}^3\beta_i)-1$ to estimate that Copilot adoption is associated with a statistically significant 18.38\% drop in MTTC.   

Contrary to the reduction in MTTC for true positives in the web portal, we find no statistically significant reduction in MTTC for alerts classified in the API or for false positive alerts.  Although our model does not pinpoint the mechanism behind the disparity, we offer conjecture in Section \ref{sec:disc}.

\subsection{Robustness}
Although our model is parsimonious, we nevertheless show that our results for true positives classified in the web portal are not sensitive to our small assumption set.  The robustness results are given in Table \ref{tab:rob}.  In general, our estimated effect ranges from over 18\% to nearly 30\%, depending on the assumption set.  All reductions are statistically significant at least at the 10\% significance level.  

\begin{table}[]
\resizebox{\textwidth}{!}{%
\begin{tabular}{|l|c|c|c|}
\hline
\textbf{Assumption} & \textbf{Effect Estimate (Reduction)} & \textbf{F-statistic} & \textbf{p-value} \\ \hline
Increase Max Classification Time to Two   Hours   & 29.78\% & 10.34 & 0.001 \\ \hline
Monthly   Fixed Effects                       & 19.91\% & 3.00     & 0.08  \\ \hline
Organization   only Standard Error Clustering & 20.31\% & 3.60   & 0.061  \\ \hline
\end{tabular}%
}
\caption{Robustness Checks}
\label{tab:rob}
\end{table}

\section{Device Policy Management}

\subsection{Data and Method}
We analyze Microsoft Intune policy conflict data from April 1st, 2024, to February 15th, 2025. We define a ``policy incident'' as a unique instance of a policy conflict. Conflict resolution often varies due to several factors, such as differences in OS type, version, and applicable policies. When policy conflicts arise, they typically appear immediately after policy deployment, and administrators must resolve these conflicts to ensure proper policy enforcement.  For each device policy conflict incident, we compute the resolution time by calculating the time difference between when the conflict was first detected and when it was successfully resolved. 
We collect incidents from Copilot adopters as well as a subset of non-adopters selected via the same propensity scoring method described in Section \ref{sec:nonmttr}.  We limit our policy conflicts to those lasting between five minutes and two days and to organizations with at least 15 resolved device policy incidents before and after adoption.  This creates a data set of 494 organizations, with 267 adopters and 227 non-adopters for a total of 207,804 policy conflict incidents. 

To estimate the impact of Copilot, we estimate a two-way fixed effects model of the form

\begin{equation}
\log(t_{ijt}) = A_i + B_{M(t)} + \beta_0 T_{it}
\end{equation}
where $t_{ijt}$ is the time to resolve device policy conflict incident $j$ for organization $i$ at time  $t$ (measured in weeks).  The variables $A_i$ and $B_{M(t)}$ are organization and time fixed effects where $M(t)$ maps week $t$ to its calendar month.  The treatment variable, $T_{it}$, indicates whether organization $i$ has adopted  Microsoft Copilot by week $t$.

\subsection{Results}
This model has only one parameter of interest and its estimate is given in Table \ref{tab:IntuneRes}.  Practically, the results indicate that Copilot adoption is associated with a statistically significant 54.34\% reduction in time to resolve a device policy conflict.  Again, we use clustered standard errors at the organization level to conduct inference. 

\begin{table}[h]
\begin{center}
\begin{tabular}{|c|c|c|c|c|}
\hline 
\textbf{Parameter} & \textbf{Estimate} & \textbf{Clustered S.E.}  & \textbf{t-statistic}& \textbf{p-value} \\
\hline \hline
$\beta_0$ & -0.783988 & 0.074809 & -10.479852& 1.069112e-25 \\
\hline
\end{tabular}
\caption{Results for GAI's impact on Policy Conflict Resolution. $R^2=0.220$}
\label{tab:IntuneRes}
\end{center}
\end{table}

\subsection{Robustness}

To ensure the reliability of our findings regarding Copilot's effect on device policy conflict resolution time, we conduct several robustness checks by varying key modeling assumptions and sample selection criteria. Table \ref{tab:intune_robust} presents the results of these robustness checks, which demonstrate that the reduction in policy conflict resolution time associated with Copilot adoption remains highly statistically significant across all specifications, with effect sizes ranging from 38.44\% to 56.01\%. The consistency of the effect direction and magnitude across multiple specifications provides strong evidence that our finding is not an artifact of specific modeling choices. When varying time fixed effects from monthly to weekly, we observe nearly identical results (54.34\% vs. 54.43\% reduction), confirming that the temporal aggregation level does not significantly affect our findings. Even when using data aggregated at the monthly level, which provides the most conservative estimate, we still observe a substantial 38.44\% reduction that remains highly significant (p=1.151e-09).

\begin{table}[h]
\centering
\resizebox{\textwidth}{!}{
\begin{tabular}{|l|c|c|c|c|c|}
\hline
\textbf{Specification} & \textbf{Coefficient} & \textbf{Clustered S.E.} & \textbf{t-statistic} & \textbf{p-value} & \textbf{\% Reduction} \\
\hline
Base model (Monthly FE) & -0.784 & 0.075 & -10.480 & 1.069e-25 & 54.34\% \\
\hline
\multicolumn{6}{|l|}{\textit{Alternative Fixed Effects:}} \\
\hline
Weekly FE & -0.786 & 0.075 & -10.540 & 5.841e-26 & 54.43\% \\
\hline
Data aggregated monthly & -0.485 & 0.080 & -6.087 & 1.151e-09 & 38.44\% \\
\hline
\multicolumn{6}{|l|}{\textit{Policy Duration Window:}} \\
\hline
10min $-$ 1day & -0.710 & 0.081 & -8.748 & $<$0.001 & 50.85\% \\
\hline
1min $-$ 2days & -0.822 & 0.070 & -11.701 & $<$0.001 & 56.01\% \\
\hline
\multicolumn{6}{|l|}{\textit{Organization Size Filtering:}} \\
\hline
Min 20 incidents/org & -0.793 & 0.076 & -10.410 & $<$0.001 & 54.76\% \\
\hline
Min 30 incidents/org & -0.802 & 0.078 & -10.264 & $<$0.001 & 55.12\% \\
\hline
\multicolumn{6}{|l|}{\textit{Robust Std Errors - No Clustering:}} \\
\hline
No clustered SEs & -0.784 & 0.004 & -223.31 & $<$0.001 & 54.34\% \\
\hline
\end{tabular}
}
\caption{Robustness of Copilot's Effect on Device Policy Conflict Resolution Time}
\label{tab:intune_robust}
\end{table}

The policy conflict duration window specifications show that our results are robust to different filtering criteria, with the effect ranging from 50.85\% (when limited to conflicts lasting 10 minutes to one day) to 56.01\% (when expanded to include conflicts lasting between one minute and two days). Similarly, increasing the minimum number of incidents per organization from 15 to 20 or 30 slightly strengthens the effect (54.76\% and 55.12\%, respectively), suggesting that organizations with more incidents may experience marginally larger benefits. The dramatic reduction in standard errors when removing clustering (from 0.075 to 0.004) illustrates the substantial correlation in the error terms within organizations and underscores the importance of our clustered standard errors approach for accurate statistical inference. Overall, these robustness checks strongly support our conclusion that Microsoft Copilot adoption is associated with substantial and statistically significant reductions in the time required to resolve device policy conflicts.


\section{Discussion and Caveats}
\label{sec:disc}
Our result indicate that Copilot adoption is associated with meaningful, statistically significant, and robust speed and accuracy productivity effects across a range of scenarios.  Copilot is associated with over a 50\% drop in the probability an incident is reopened, suggesting that Copilot leads to better initial decisions for incident management.  Furthermore, the number of alerts per incident decreases by 23\%, suggesting that analysts are resolving incidents earlier in the kill chain with less information.   

For DLP, our results show that Copilot is robustly associated with a significant reduction in MTTC for true positive alerts classified in the web portal.  The lack of effect for the API is not surprising since API processing likely captures a significant number of alerts triaged by automation tools where Copilot is not yet designed to assist.  

The lack of an effect for false positives is interesting.  One reason may be that there are correlations between false positives.  That is, learning about one false positive can facilitate discovery of other false positives (possibly due to a misconfigured policy).  True positives on the other hand are unique to specific attacks and malicious behavior.  Therefore, while Copilot may help uncover an initial false positive, it doesn’t impact the classification time for similar alerts, thus limiting the number of alerts where Copilot is effective.  On the other hand, malicious behavior is unique and thus Copilot is applicable to all true positive alerts.  Nevertheless, this is an unproven conjecture and a possible direction for future work.

For device policy conflicts, the results show that Copilot is robustly associated with a 54\% reduction in MTTR. Copilot's policy conflict feature directly addresses a common problem faced by IT administrators when designing and deploying new policies. It summarizes which devices and settings are in conflict, saving the time it would otherwise take to go through each individually. 

From a certain perspective, it is surprising that GAI offers such significant productivity gains across scenarios that have long been central to security operations. The processes and technology stacks in modern security operations are already highly optimized. To provide the substantial incremental improvements that we and others measure, GAI must fill a previously unfilled gap.  

\subsection{Caveats and Future Work}
Although we use the tools of causal inference, it is impossible to rule out selection into treatment. For example, an organization may adopt Copilot because the information security department received a budget expansion. So, even though Copilot may contribute to productivity gains, the co-occurrence of unobserved factors like an increased number of analysts, other new software licenses, and additional trainings make it impossible to isolate the impact of Copilot on productivity using observed telemetry alone.

 Moreover, it is even likely that those that are willing to pay for Copilot are the ones that would most benefit from Copilot. In other words, when organizations perform a cost/benefit analysis on the decision to adopt Copilot, only those with the highest benefit would adopt Copilot, since all organizations face the same sticker price. If so, this would mean that even perfectly identified causal estimates would overstate the average productivity impact of Copilot for \textit{non-adopters}.

Considering these limitations, our results provide evidence for the causal impact of Copilot on productivity. However, short of performing a randomized control trial on live organization operations or discovering a natural experiment that mimics randomized selection, the associations uncovered in this study are among the most promising signals supporting the positive effect of GAI tools (like Copilot) on the productivity of security operations and endpoint management. Furthermore, the results are statistically significant and robust, so it is clear that those that adopted Copilot observed a productivity gain.

\bibliographystyle{plain}
\bibliography{refs}

\end{document}